\documentstyle[12pt]{article}
\newcommand\pubnumber{ TAUP-2291-95}
\newcommand\pubdate{Sept, 1995}
\newcommand\pubtype{T}

\def\Title#1{\begin{center} {\Large #1 } \end{center}}
\def\Author#1{\begin{center}{ \sc #1} \end{center}}
\def\Address#1{\begin{center}{ \it #1} \end{center}}
\def\andauth{\begin{center}{and} \end{center}}
\def\doeack{\footnote{Work supported by the Department of Energy,
                     contract DE--AC03--76SF00515.}}
\def\SLAC{Stanford Linear Accelerator Center\\
    Stanford University, Stanford, California 94309 USA}
\newcommand{\journal}[4]{{\sl #1}\ {\bf #2}, #3\ (#4)}
\newcommand\pubblock{\rightline{\begin{tabular}{l} \pubnumber\\
         \pubdate \\ (\pubtype) \end{tabular}}}
\newenvironment{Abstract}{\begin{quotation} \begin{center}
                       ABSTRACT
     \end{center}\bigskip  }{\end{quotation}}

\def\beq{\begin{equation}}
\def\eeq#1{\label{#1}\end{equation}}
\def\eeqn{\end{equation}}
\def\beqa{\begin{eqnarray}}
\def\eeqa#1{\label{#1}\end{eqnarray}}
\def\eeqan{\end{eqnarray}}
\def\CR{\nonumber \\ }
\def\leqn#1{(\ref{#1})}

\def\Acknowledgements{\bigskip  \bigskip {\begin{center} \begin{large}
             \bf ACKNOWLEDGEMENTS \end{large}\end{center}}}
\def\bar#1{\overline{#1}}
\def\Dslash{\not{\hbox{\kern-4pt $D$}}}
\def\notR{\not{\hbox{\kern-4pt $R$}}}
\def\Nf{{N_f}}
\def\Nc{{N_c}}
\def\Qb{{\bar{Q}}}

\def\Bb{{\bar{B}}}
\def\bq{{\bar{q}}}
\def\dT{{T^{\dagger}}}

\def\half{{1\over 2}}

\def\L{{\cal L}}
\def\M{{\cal M}}
\def\mQ{{m_Q^2}}
\def\mg{{m_g}}

\def\Mg{{M_g}}
\def\tr{{\rm tr}}

\def\VEV#1{\left\langle{ #1} \right\rangle}
\def\ra{\rightarrow}

\def\hc{{\rm h.c.}}
\def\coback{\footnote{Work supported in part by the Israel Academy of Science}}
\def\ofack{\footnote{Work supported in part by the Clore Scholars Programme}}
\def\telack{\footnote{Work supported in part by the US-Israel Binational
Science Foundation and by GIF -- the German-Israeli Foundation for
Scientific Research}}
\def\shimack{\footnote{On leave of absence from the School of Physics,
Raymond and Beverly Sackler Faculty of Exact Sciences, Tel-Aviv University}}
\def\telaviv{School of Physics and Astronomy\\ Beverly and Raymond Sackler
Faculty of Exact Sciences\\ Tel Aviv University,
    Ramat Aviv, Tel Aviv 69978, Israel}
\def\CERN{CERN, Geneva, Switzerland}

\begin{document}

\begin{titlepage}
\pubblock

\Title{ Duality and other
Exotic Gauge Dynamics \\  in Softly Broken Supersymmetric QCD}

\Author{Ofer Aharony\ofack$^,$\coback$^,$\telack, Jacob Sonnenschein$^{2,3}$}
\Address{\telaviv}
\Author{Michael E. Peskin\doeack}
\Address{\SLAC}
\andauth
\Author{Shimon Yankielowicz$^{3,}$\shimack}
\Address{\CERN}
\begin{Abstract}

We analyze the theory of softly broken supersymmetric $QCD$.
Exotic behavior like spontaneously broken baryon number,
massless composite fermions and
Seiberg's  duality seems to persist also in the
 presence of (small) soft supersymmetry breaking.
 We argue that certain, specially tailored, lattice
simulations may be able to detect the novel phenomena. Most of the exotic
behavior does not survive the decoupling limit of large SUSY breaking
parameters.

\it{Based on a talk presented  by J. S in SUSY 95.}
\end{Abstract}
\end{titlepage}
\section{Introduction and summary}

\begin{description}
\item[*]
(1) We study  multi-flavored four dimensional  QCD  by adding soft
supersymmetry
breaking (SSB) terms to supersymmetric QCD (SQCD)
\item[*]
(2) Certain exotic phenomena discovered in SQCD persist also in the regime of
small, compared to $\Lambda_{QCD}$, SSB parameters.
\item[*]
(3)  The exotic behavior includes (i)  a baryon number violating vacuum, (ii)
massless composite fermions and (iii)  Seiberg's duality\cite{NatiDual}.
\item[*]
(4)   In the decoupling limit most of  the
``exotic" phenomena disappear.
\item[*]
(5)  We argue that all the exotic behavior
    of the region  close to the SUSY limit should be detected in
lattice simulations.
\end{description}
\section{On the effective action of SQCD }
$SU(\Nc)$ SQCD  contains a
gauge superfield $W_\alpha$,
 a chiral quark and an anti quark superfields
$ Q^i_a, \Qb^a_i$
where $i= 1, \ldots, \Nf$ is a flavor index and
 $a = 1, \ldots, \Nc$\footnote{ We take here $\Nc\ge 3$. The special case
of $\Nc=2$ is discussed in ref.\cite{us}.}
 is a color index. These superfields transform under the quantum global
symmetry
\beq
   SU(N_f)_L \times SU(N_f)_R \times U(1)_B \times U(1)_R ,
\eeq{globalsymm}
 in the following way
 \begin{center}
\begin{tabular}{|c|c|c|c|c|c|}\hline
 $ ~~$~~~  & ~~ $SU(N_c)$~~~  & ~~ $SU(N_f)_L$ ~~ & ~~$SU(N_f)_R$ ~~ &~~
$U(1)_B$ ~~ & ~~$U(1)_R$~~ \\
\hline
$Q$ & $ N_c$ & $ N_f$  & 1 & ${1\over N_c}$ &  $ (N_f - N_c)/N_f$ \\
$\bar{Q}$& $ \bar N_c$ & 1 & $\bar{N}_f$ & ${-1\over N_c}$ &$ (N_f -
N_c)/N_f$\\
$W_\alpha$ & $ N_c^2-1$ & $ 1$  & 1 & $0$ &  $ 1$ \\
\hline
\end{tabular}
\end{center}
The low energy effective action of SQCD  is expressed,
 for small values of $\Nf$, in terms of
 the meson superfield
\beq
      T^i{}_j =  Q^i \cdot \Qb_j.
\eeq{mesons}
Beginning at $\Nf = \Nc$, $S_{eff}$ is built  also from  chiral baryon and
anti-baryon  superfields. The baryon
\beq
     B_{i_1 \cdots i_{(\Nf - \Nc)}} =   \epsilon^{a_1 \cdots a_\Nc}
\epsilon_{j_1 \cdots j_\Nc i_1 \cdots i_{(\Nf - \Nc)}} Q^{j_1}_{a_1} \cdots
            Q^{j_\Nc}_{a_\Nc}  ,
\eeq{baryonop}
 transforms in the $(\Nf - \Nc)$-index antisymmetric
tensor representation of $SU(\Nf)_L$,
and, similarly, an antibaryon chiral superfield $\Bb^{i_1 \cdots
i_{(\Nf-\Nc)}}$ is built from $\Nc$ powers of the field $\Qb$.

 Using the gauge supermultiplet,
it is possible to build the  (massive) ``glue ball" chiral superfield
\beq
     S = - \tr\bigl[ W^{\alpha } W_\alpha\bigr] = \tr[ \lambda \cdot
     \lambda]
                  + \cdots \ .
\eeq{Sdef}
 The K\"ahler potential
 which determines the kinetic energy terms of the fields $T$, $B$,
and $\Bb$ was  hypothesized in the work of Masiero and
Veneziano \cite{MV,MPRV} to take the following form
\beq
    K[T,B,\Bb] =   A_T \tr \bigl[ T^\dagger T \bigr] +
         A_B   \bigl( B^\dagger B + \Bb^\dagger \Bb \bigr) .
\eeq{cKahler}
Our main results will rely on weaker assumptions about the K\"ahler
potential, in particular, that it is nonsingular on the space of
supersymmetric vacuum states.
 However, we will support our general
remarks by explicit calculations using this simple model.
We expect \leqn{cKahler} to be the correct form of the K\"ahler
potential near the origin of moduli space, in the cases for which the
mesons and baryons give an effective infrared  description of the theory.
This is not the case for $\Nc>\Nf$\cite{uss}.

\subsection{Soft Supersymmetry Breaking}
 We  break supersymmetry by adding mass terms for the
squarks and gaugino fields,
\beq
\Delta \L =   - \mQ \Bigl( \bigl| Q \bigr|^2 +  \bigl|\Qb \bigr|^2\Bigr)
                          + \bigl( \mg S  + {\rm h.c.}\bigr),
\eeq{softbr}
where, in \leqn{softbr}, $Q$, $\Qb$, and $S$ are the scalar component fields of
the superfields.  The squark  mass term is the unique soft supersymmetry
breaking term which does not break any of the global symmetries
\leqn{globalsymm} of the original model.  The gaugino mass term breaks only
the $U(1)_R$ symmetry, and thus breaks the global symmetry of the
supersymmetric model down to that of ordinary QCD  with $\Nf$
massless flavors.
Actually, there are two consistent possibilities for   SSB  theories
which we denote by $R$ and $\notR$
\beqa
R:& \mQ\ne 0\   \mg=0 \ \   \Longrightarrow_{\mQ\rightarrow\infty}
 {\it standard}\  QCD\  with\ N_f\
{\it quarks \ +\ adjoint\  fermions}\CR
\notR:& \mQ\ne 0 \  \mg\ne0\ \
\Longrightarrow_{\mQ,\mg\rightarrow\infty}
{\it standard}\  QCD\  with\ N_f\ {\it quarks} \nonumber
\eeqan
Since we will be working in the language of the low-energy effective
Lagrangian, we must ask how the supersymmetry breaking term
\leqn{softbr} shows up in this Lagrangian.  To work this out, rewrite
\leqn{softbr} in the superfield form
\beq
    \Delta\L =  \int d^4\theta  M_Q \bigl( Q^\dagger e^V Q
 +\Qb^\dagger e^{-V^T} \Qb\bigr)
                    + \int d^2\theta M_g S + {\rm h.c.}
\eeq{softbrsf}
where $M_Q$ is a vector superfield whose $D$ component equals $(-m_Q^2)$ and
$M_g$ is a chiral superfield whose $F$ component equals $m_g$.
It is straightforward to see that these superfields are
gauge-invariant and neutral under all of the global symmetries.

 The
effective Lagrangian description of $\Delta\L$ for $N_f \leq N_c+1$
is then given by writing
the most general Lagrangian built from $T$, $B$, $\Bb$ and a fixed
number of factors of $M_Q$ and $M_g$.  The supersymmetry breaking terms
have an ambiguity related to that of the K\"ahler potential, because many
possible invariant structures can be built from $T$
$B$, and  $\Bb$.  In our explicit calculations,
we will assume that the coefficient of $M_Q$ is quadratic in these
fields. This assumption holds near the origin of moduli
space.  Then the first order soft supersymmetry breaking terms in the
effective Lagrangian are
\beqa
 \Delta\L & = \int d^4 \theta  \Bigl(
    B_T M_Q \tr\bigl[ T^\dagger   T\bigr]  + B_B M_Q \big\{
     B^\dagger B + \Bb^\dagger \Bb\bigr\} + \bar{M_g} \M(T,B,\Bb) + \hc
      \Bigr) \CR
         & \quad + \int d^2 \theta  M_g \VEV{S} + \hc ,
\eeqa{SSoft}
{\it clarifications and remarks}:

(i)  $\M(T,B,\Bb)$ is a function of the effective Lagrangian
superfields which is neutral under the global symmetries.

(ii) The quantity $\VEV{S}$ in \leqn{SSoft} should be
a combination of the effective Lagrangian
chiral superfields which has the quantum numbers of $S$. In general,
this condition restricts that function to be proportional to the
expectation value of $S$ as determined from the effective Lagrangian
of refs. \cite{VY,TVY} which includes $S$ as a basic field.

(iii) The squark mass terms in \leqn{SSoft} are not the most general terms that
can be written down.  Higher order
terms in the fields, suppressed by powers of $\Lambda$, may appear.
However, we expect \leqn{SSoft} to be approximately true near the origin
of moduli space $T=B={\bar B}=0$.

(iv)
We have assumed throughout this work that the coefficients $B_B$ and $B_T$
are positive. This is not an innocent assumption since had they been negative
chiral symmetry breaking would  have taken place instead of several of the
exotic  features.
At present we do not have a proof of the positivity of those coefficients,
but the picture we derive assuming that they are positive seems to be
consistent.

(v) The ratio of coefficients $B_B/B_T$ will be important to our later
analysis, but this ratio
cannot be determined from the effective Lagrangian viewpoint.  At best,
we can argue naively that the coefficient of the mass term of a
composite field should be roughly proportional to the sum of the
coefficients of the mass terms of the constituents.  This would give
the relation $ B_B \approx {\Nc\over 2} B_T$
which the reader might take as qualitative guidance.

(vi) It is important for our analysis that the behavior of the theory is
non--singular when adding the squark and gluino masses, i.e. that no new
non--perturbative effects occur. In general it is not possible to prove
this in non--supersymmetric theories, but a proof of this is possible in
softly broken supersymmetric theories, when the soft breaking can be viewed
as arising via  spontaneous breaking of supersymmetry.\cite{Evans}



\section{ Baryon number violation ($\Nf=\Nc$)}
The cases  of $\Nc>\Nf$ are discussed in ref.\cite{us}, here we start with
 the case $\Nf = \Nc$.  The
low-energy effective Lagrangian of the supersymmetric limit contains
both meson and baryon superfields.
The quantum numbers of the $T^i{}_j$, $B$ and  $\Bb$ are summarized in the
following table.

\begin{center}
\begin{tabular}{|c|c|c|c|c|c|}\hline
 $ ~~$~~~  & ~~ $SU(N_c)$~~~  & ~~ $SU(N_f)_L$ ~~ & ~~$SU(N_f)_R$ ~~ &~~
$U(1)_B$ ~~ & ~~$U(1)_R$~~ \\
\hline
$T^i_j$ & $ 1$ & $ N_f$  & $\bar N_f$ & $0$ &  $ 0$ \\
$B$ & $ 1$ & $ 1$  & $1$ & $1$ &  $ 0$ \\
$\bar B$ & $ 1$ & $ 1$  & $1$ & $-1$ &  $ 0$ \\
\hline
\end{tabular}
\end{center}

This model has a   manifold of quantum mechanical supersymmetric
ground states, in which the meson and baryon fields satisfy the
relation (in units where $\Lambda = 1$)
\beq
      \det T  -   B \Bb =   1 \ .
\eeq{Sconstraint}
Several  forms for the superpotential are consistent with this relation.
The $S$-dependent superpotential, for example, has the form
    $ W  =   S \log \bigl(\det T - B\Bb \bigr)$ .
Note that this superpotential leads to conditions for a
supersymmetric vacuum state which imply not only \leqn{Sconstraint}
but also the constraint $\VEV{S} = 0$, so that the $U(1)_R$ symmetry is
not spontaneously broken.

The presence of a  manifold of degenerate vacuum states not related
by a global symmetry is necessarily accidental unless it is
 a consequence of supersymmetry.  Thus, any such
degeneracy should be broken as soon as supersymmetry breaking terms are
added to the Lagrangian.  To first order, this is the main effect of the
soft supersymmetry breaking perturbation.  To analyze this effect, we
should restrict our attention to the values of
 $T$, $B$, and $\Bb$ obeying the constraint
\leqn{Sconstraint}, for which the vacuum energy vanishes in the
supersymmetric limit, and study the behavior of the supersymmetry
breaking potential over this space.

For the {\it  $R$ models}, the soft supersymmetry breaking terms
\leqn{SSoft} lead to the potential
\beq
  \Delta V =  B_T \mQ \tr \bigl[ T^\dagger T \bigr]  + B_B \mQ \bigl(
                          B^\dagger B + \Bb^\dagger \Bb \bigr) .
\eeq{delVwB}
Using $SU(\Nf)\times SU(\Nf)$, we can diagonalize $T$ to complex
eigenvalues $t_i$.  Parameterize the baryon fields as
$B = x b \ ,\ \Bb = - {1\over x}b$
with $x$ and $b$ complex. This potential has   three types of

\centerline{\it stationary points }
\centerline{$\swarrow \ \ \ \ \ \  \Downarrow\ \ \  \ \ \ \ \  \searrow \ \ \ $
}   \begin{center}
\begin{tabular}{c| c| c}
$b=0$~ $\forall_i\ |t_i|=1$ & $T\ne 0 $, $b\ne 0$ ~ ~ & $T=0$, $b=\pm 1$\\
  & $|t_i|^{(\Nf-2)} = (B_T/B_B)$  &\\
& unstable  &
for\  $B_B > (\Nf/2)B_T$\\
& &  globally\  stable \\
\end{tabular}
\end{center}
with $x=1$ in all cases.
The method of effective Lagrangians cannot tell us which of the two
vacuum states at $b=0$ and $T=0$ is the preferred one.  This depends on
the ratio  $B_B/B_T$, which is a phenomenological input to the
effective Lagrangian analysis.
The vacuum at
$T=0$ is locally stable if $B_B > B_T$ and is globally stable if
$B_B > (\Nf/2)B_T$.   Our naive estimate puts the theory just at
the boundary at which the two vacuum states have equal energy.

The structure of  the {\bf  spectrum} in the $b=0$ and $T=0$ cases is
summarized as follows (for details see \cite{us}).
\begin{center}
\begin{tabular}{|c|c|c|}\hline
& $b=0$\  vacuum  & $T=0$\  vacuum \\ \hline
Symmetry& $ SU(N_f)_D \times U(1)_B \times U(1)_R$ &
 $ SU(N_f)_L \times SU(N_f)_R  \times U(1)_R$\\
parametrize& $ B = b + c \ ,  \qquad      \Bb =   - (b -c ) $ &
 $B = (1 + b + c ) \ ,   \Bb = - (1 + b - c ) \ $\\
constraint & $t_V + i t_A = -  \sqrt{{2\over \Nf}} (b^2 - c^2) \ $&
     $ b = - \half \det T  + \half c^2 \ $\\
bosons&  $t_{A_I}$- NG bosons $t_A$ massive & $Im(c)-NG boson$\\
    & $m^2_{Vi}  =    {2\over A_T} B_T \mQ$  &
     $ m_T^2 = {B_T\over A_T} \mQ$\\
 & $m^2_-     =  {2\over A_B} (B_B - B_T) \mQ$ &
$m_{cR}^2 = {B_B\over A_B} \mQ$
\\
 & $m^2_+     =  {2\over A_B} (B_B + B_T) \mQ $ & \\
fermions& $m_{\psi_T}=\ m_{\psi_B}= m_{\psi_\Bb}=0$ &
 $m_{\psi_T}=\ m_{\psi_B}= m_{\psi_\Bb}=0$ \\
 \hline
\end{tabular}
\end{center}

Notice  that since  \leqn{Sconstraint} is a superfield constraint, it also
removes one fermion from the theory, specifically, the fermionic
partner of $det[T]$ or of $B+\Bb$.

For the {\it $\notR$ models} with small $\mg$ the qualitative picture remains
the
same.
  We showed earlier that
 the superpotential  implies that,
in the manifold of
supersymmetric vacuum states about which we are perturbing, $\VEV{S} =
0$.  Thus, the superpotential term proportional to $\Mg$ does not
contribute to the vacuum energy.  More generally, since $\Mg$, $T$,
$B$, and $\Bb$ are all invariant under $U(1)_R$, while a superpotential
has $R$ charge 2, this term does not contribute to the superpotential
to any order in $\mg$.  There are possible K\"ahler potential terms
involving $\Mg$.
However, near the vacuum with $b= 0$, these will be polynomials in
$B$ and $\Bb$ of order at least 2, and near the vacuum with $T=0$ they
will be polynomials in $T$ of order at least 2.  Thus, these terms will
not affect the presence of stationary points of the vacuum energy at
these positions in the field space.
Examples of $D$-term terms that can be induced are
\beq
   \Delta \L = \int d^4 \theta \bar{\Mg} \bigl( C_T \det T + C_B
   B\Bb\bigr)  + \hc \ ,
\eeq{FirstCtry}
where $C_T$ and $C_B$ are some constants.  If one begins from the
effective Lagrangian including $S$, with the canonical superpotential
and K\"ahler terms,
\beq
   \L = \int d^4 \theta  S^* S + \int d^2\theta \bigl( S\log(\det T -
   B\Bb) + \Mg S \bigr) + \hc \ ,
\eeq{fullSLag}
and integrates out $S$, one finds a breaking term
\beq
   \Delta \L = \int d^4 \theta \bar{\Mg} \log(\det T - B\Bb)  + \hc ,
\eeq{SecondCtry}
which gives qualitatively similar results.  In the following
discussion, we will work with \leqn{FirstCtry}. Once the $U(1)_R$ is broken
it does not protect fermions from acquring mass.
Around $b=0$
 the flavor adjoint and baryonic fermions,
acquire the following  masses
     $m_{\psi i }  =  \half {C_T\over A_T} \mg$ and
     $m_{\psi B}  =  {C_B \over A_B} \mg$
and no zero-mass fermions remain.
Around the $T=0$ vacuum $\psi_{T_I}$ and $\psi_{T}$ remain massless  but
the baryonic fermion aquires  a mass
      $m_{\psi B}  =  {2C_B \over A_B} \mg$.

\subsection{Toward the Decoupling Limit}
The
transition from the region of weak supersymmetry breaking
 to the decoupling limit $\mQ, \mg \ra \infty$  is very different in the
four models  of $R$ and $\notR$ with vacua at $b=0$ and $T=0$.

{\bf (i) $\notR$\  b=0}-
In this case (like in the cases of $\Nc>\Nf$)
the global group $SU(\Nf)\times SU(\Nf) \times U(1)_B$ is
broken spontaneously to $SU(\Nf)\times U(1)_B$,
leaving no massless particles except
for the required Goldstone bosons.  It is reasonable to expect that
 there is a smooth
transition from the situation of weak supersymmetry breaking
 to the decoupling limit $\mQ, \mg \ra \infty$.
 In QCD, chiral symmetry breaking is characterized by
a nonzero vacuum expectation value of the quark-antiquark bilinear,
 $\psi_Q^i\psi_{\Qb j}$ in our present notation.  In the language of
the supersymmetric effective Lagrangian, this operator is a part of the
$F$ term of the superfield $T^i{}_j$.  The expectation value of this term
may easily be found\cite{us} to be proportional to
 $m_Q^{N_c\over{2N_c-N_f}}$.
Thus, the $F$ term of $T$ does obtain an expectation value in the
vacuum state that we have found. This expectation value naturally
becomes a nonzero expectation value for the quark bilinear in the
decoupling limit. As $m_Q$ increases, the quark bilinear
becomes larger while the squark bilinear becomes smaller, in exact
accord with our expectations.

{\bf (ii) $R$\  b=0}-
   As $\mQ$ is taken to
infinity, the squarks decouple, and the model becomes a purely fermionic
$QCD$ theory with $\Nf$ quark flavors plus one fermion flavor
in the adjoint representation of the gauge group.
For small values of the
supersymmetry breaking mass $\mQ$,  this vacuum contains massless
fermions corresponding to the fermionic components of the superfields
$T$, $B$, and $\Bb$.  We might think of these as being built out of
scalars, with one squark replaced by a quark to give the composite spin
$\half$.  But it is also possible to build
composite bound state with the same
quantum numbers purely  out of fermions, by replacing
\beq
            Q^i \ra  \lambda^\alpha \psi_{Q \alpha}^i\ , \qquad
               \Qb_j \ra  \lambda^\alpha \psi_{\Qb \alpha j} \ ,
\eeq{Qreplace}
where $\alpha$ is a two-component spinor index and the gauge
indices are implicit.
Notice that this combination has the same quantum numbers as the
squark, including zero $R$ charge.  Then, for
example, the fermion created by $T^i{}_j$ could be constructed as
\beq
       \psi_{T\alpha} {}^i{}_j \ra  \psi_{Q \alpha}^i \lambda^\beta
                   \psi_{\Qb\beta  j} \ .
\eeq{fmesonform}
With this replacement, the composite fermions are built only out of
constituents which remain massless as the squarks are decoupled.  Thus,
it is apriori reasonable that the $b = 0$ vacuum of the $R$ model
could go smoothly into a vacuum of the  purely fermionic
QCD theory described above.  This vacuum would have broken
$SU(\Nf)\times SU(\Nf)$ but unbroken chiral $U(1)_R$, zero values for
the vacuum expectation values of quark-antiquark bilinears,
massless composite fermions in the adjoint representation
of flavor $SU(\Nf)$, and massless baryons.
We will refer to this
scenario as `option 1'.

The other possibility for this model is that, after the squarks
decouple, the gluino fields pair-condense,
in a second-order phase transition at some value of $\mQ$,
and the nonzero value of
the condensate $\VEV{\lambda\cdot \lambda}$
spontaneously breaks $U(1)_R$.  In this
case, the physics would revert to the usual symmetry-breaking pattern
of QCD, and the composite fermions would become massive.
The gluino condensate would make itself felt only by providing an extra
$SU(\Nf)$-singlet  Goldstone boson.  We will refer to this scenario as
`option 2'.

{\bf (iii) $\notR$\  T=0}-
In this model,  in the supersymmetric limit,
the massless composite fermions belong to the
$(\Nf,\bar{\Nf})$ representation of an unbroken flavor group $SU(\Nf)\times
SU(\Nf)$. There are no constraints from the $U(1)_R$ symmetry, which is
explicitly broken, or from baryon number, which is spontaneously
broken.  With this freedom,
can we build these fermionic composites out of fields that survive in
the decoupling limit $\mQ, \mg \ra \infty$?  For $\Nc$ even, it is
impossible, because the only constituents available are the quarks
$\psi_Q^i$, $\psi_{\Qb j}$, and gauge-invariant states must contain
an even number of these.  For $\Nc$ odd,
however, it is possible to build
composites with the correct quantum numbers, as follows:
\beq
       \psi_{T\alpha} {}^i{}_j \ra  \epsilon^{ab \cdots d}\,
      \psi_{Q\alpha  a}^i\,  \epsilon_{jk\cdots \ell}
         \bar{\psi}^{\dot{\beta} k}_{\Qb b } \cdots
          \bar{\psi}^{\ell}_{\Qb \dot{\gamma} d}  \ ,
\eeq{Tmesonform}
where $\bar\psi_{\Qb}$ is the right-handed fermion field in
$(\Qb)^*$. The $(\Nc -1)$ right-handed fermion fields must be
contracted into a Lorentz scalar combination.  For the case $\Nf = \Nc
= 3$, eight of the nine fermions in \leqn{Tmesonform} have the
quantum numbers of the baryon octet in QCD.

However, in this case, there are two compelling arguments that the
spectrum which we find cannot survive to the decoupling limit.  In the
limit $\mQ \ra \infty$, even without  introducing $\mg$, we have a
vectorlike gauge theory of fermions.  For such theories, the QCD
inequalities of Weingarten \cite{Weingarten} and Vafa and Witten
\cite{VafaWitten} apply.  In \cite{us}  we use Weingarten's method
to prove that, in the decoupling limit, flavor nonsinglet composite
fermions must be heavier than the pions, which are massive in the $T=0$
vacuum.  Alternatively, we can apply the theorem of Vafa and Witten in
the decoupling limit to show that
vectorlike global symmetries, in particular, baryon number, cannot
be spontaneously broken.

By either argument, the $T=0$ vacuum state
must disappear via a second order phase transition at a finite value of
$\mQ$.
  Most likely, this vacuum becomes
locally unstable  with respect to a decrease in
the expectation value of $b$, driving the theory back to the
more familiar vacuum  at $b=0$.

{\bf (iv) $R$\  T=0}-
In the supersymmetric limit  we have
massless fermions in the following representations
$(\Nf, \bar{\Nf},-1) +
(1,1,-1)$  of the unbroken symmetry group.
corresponding to the fermions in $T$ and a linear combination of
the fermions in $B$ and $\Bb$.  Both
multiplets are necessary to satisfy the anomaly conditions involving
$U(1)_R$.
  The arguments that we have just
presented for the $T=0$ vacuum in the
$\notR$ models apply equally well to the $R$ case.  Again, we must
 have a second-order transition,  probably with an
instability to the $b =0$ vacuum.  There are then two possible
endpoints, depending on which option is chosen for the $b=0$ vacuum.
If the option 1 for the $b=0$
vacuum is correct, it is not necessary that $U(1)_R$ be spontaneously broken
in this transition.

\section{ No $\chi SB$, massless fermions ($\Nf=\Nc+1$)}

In the case of $\Nf = \Nc+1$,  the low-energy
effective Lagrangian  in the supersymmetric limit is expressed  in terms  of
the
baryon, anti-baryon and meson superfields which
transform under  the global symmetry as follows
\leqn{globalsymm}
\beq
B\ : \  (\bar\Nf,1)_{1,1-{1\over \Nf}}\qquad \bar B:\ (1,\Nf)_{-1,1-{1\over
\Nf}}
 \qquad  T :\  (\Nf,\bar\Nf)_{0,{2\over \Nf}}
\eeq{NfNc1sym}
where the second subscript is the
$R$ charge of the scalar component of the superfield.
In the supersymmetric theory
the low energy effective theory is described (at least near the origin
of moduli space) by
the K\"ahler potential given by \leqn{cKahler},
and by the following  superpotential \cite{NatiMod}:
\beq
W = B_i T^i_{j}\bar B^{j} - \det T.
\eeq{W5}
The supersymmetric  vacuum is, thus,   described by a moduli
space characterized by
\beq
B_i T^i_{j} =\ 0\qquad  T^j_{i} {\bar B}^{ i}
 =\ 0\qquad {1\over N_c!}\epsilon_{i_1,...,i_{N_f}}
\epsilon^{ j_1,..., j_{N_f}} T^{i_1}_{ j_1}...
T^{i_{N_c}}_{ j_{N_c}} - B_{i_{N_f}} \bar B^{ j_{N_f}}=0.
\eeq{modspa}
As was argued in \cite{NatiMod}, these equations
correctly
describe the moduli space of vacuum states in the full
quantum theory.
At     the origin  of the moduli space,
 $<T> = <B>=<\Bb>=0$,  where the full global
symmetry \leqn{globalsymm} remains unbroken,
there  is a further consistency check for the low energy behavior.
The  fermionic components of the low-energy superfields \leqn{NfNc1sym}
match the
global anomalies of the underlying theory.

When we break supersymmetry by squark and gluino masses, we add to the
effective Lagrangian the mass terms for $T$, $B$ and $\Bb$ indicated in
\leqn{SSoft}.  Since we are adding terms to the potential which are
positive  and vanish at the origin of moduli space, it is
obvious that the origin becomes the only vacuum state of the theory.
All of the scalar particles in the effective theory obtain mass
terms proportional to $B_Tm_Q^2$ or $B_B m_Q^2$.

Though all of the scalars obtain mass, all of the fermions remain
massless.  The superpotential \leqn{W5} is a least
cubic in the fields, thus, any mass term derived from this superpotential
vanishes at the origin.  Similarly, in the $\notR$ case,
the $M_g$ term in \leqn{SSoft}
requires a function of $T$, $B$, and $\Bb$ which is neutral with
respect to the global group; the only such functions, quadratic in
fields,  are $\tr[T^\dagger T]$, $B^\dagger B$, and $\Bb^\dagger \Bb$,
and these do not give rise to  fermion masses when multiplied with $M_g$.
In fact, it is required that no fermions should obtain mass, since the
full multiplet of fermions in $T$, $B$, and $\Bb$ is needed to
satisfy the `t Hooft anomaly conditions for the global symmetry group
$SU(\Nf)\times SU(\Nf) \times U(1)_B\times U(1)_R$.

\section{ Seiberg's duality in SSB models   ($\Nf \geq \Nc+2$)}

No solution of the 't Hooft anomaly matching conditions for
SQCD
involving gauge invariant
bound states is known for $N_f > (\Nc +1)$.
 However, Seiberg has
suggested a compelling solution to these constraints in terms of
a novel  type of {\it duality} in which  new
gauge degrees of freedom  are dual to the original quarks and
gluons \cite{NatiDual}.  The pair of dual  theories are

\begin{center}
\begin{tabular}{|c| c| }\hline
{\bf ``Electric theory"} ~~~~~~~~~~~~~~~& {\bf ``Magnetic theory"}
{}~~~~~~~~~~~~~~~\\
\hline
$SU(\Nc)$\ gauge\ theory & $SU(\Nf-\Nc)$\ gauge\ theory\\
$a = 1, \ldots, \Nc$\ & $\hat a = 1, \ldots,(\Nf-\Nc)$\\
$ Q^i_{a},
(\Nf,\ 1,\ {1\over N_c},\  (N_f - N_c)/N_f)$ &
$ q_i^{\hat a},
(\bar \Nf,\ 1,\ {1\over \Nf-N_c},\   N_c/N_f)$ \\
$ \Qb^a_i
(1,\ \bar\Nf,\ -{1\over N_c},\  (N_f - N_c)/N_f)$ &
$ \bar q^i_{\hat a},
(1,\  \Nf,\ -{1\over \Nf-N_c},\   N_c/N_f$ )\\
\ &\ \ \ \ $ T^i_j (\Nf,\ \bar\Nf,\ 0,\ 2(N_f - N_c)/N_f$) \\
$W=0$\ & $W = T^i{}_j q_i^{\hat a} \bq^j_{\hat a}$\\ \hline
\end{tabular}
\end{center}
On top of the fact that these theories have the same 't Hooft anomalies, they
were shown to be equivalent in the infrared also by
analyzing their ``chiral rings", flat directions and mass perturbations.
In the supersymmetric limit of the  magnetic theory
  the scalar potential has  a
moduli space of vacua, which includes the point $<T> =<q> =< \bq> = 0$ at which
the chiral symmetry is unbroken \cite{NatiDual}.

Now add squark masses to the theory. Their effect should be seen in the
effective Lagrangian. In case that the latter is given by the dual theory
 we should  add to the effective Lagrangian of the dual theory the term
\beq
\Delta V = B_T m_Q^2 \tr(\dT T) + B_q m_Q^2 (|q|^2 + |\bq|^2),
\eeq{deltav}
at least near the origin of moduli space. After we add this
perturbation, the only minimum of the potential is at $<T> = <q> =
<\bq> = 0$.
Thus, adding a squark mass leaves the theory in the phase in which the
chiral symmetry is unbroken. All scalars get masses (originating only from
$\Delta V$, since the original scalar potential is quartic in the fields),
while all fermions remain massless.
As in the original supersymmetric theory, this complement of massless
fermions has just the right quantum numbers to satisfy the `t Hooft
anomaly conditions for completely unbroken chiral symmetry.

The glueball operator $\tr(W_{\alpha}^2)$ is identified (up to a sign)
between the original and the dual theory \cite{NatiIntDual}.
Thus, to leading order in $m_g$,
a gluino mass in the original theory is just equal to a gluino mass in the
dual theory. Adding this term breaks the $U(1)_R$ symmetry, but the
$SU(N_f)\times SU(N_f)$ global symmetry still remains and protects the
dual quarks from getting a mass. Thus, we find the same spectrum in the $R$
and $\notR$ cases, except that in the latter case the dual gluino,
which can be an asymptotic particle, becomes massive.

Let us discuss now the infrared description of the theory.
For small $m_Q$ (and small
 $m_g$, in the $\notR$ case) the situation is schematically described in the
following diagram
\begin{center}
\begin{tabular}{c c c c c c c c }
{electric theory}&  &  &   & {\bf AS\ free} *   & : &\ {\bf IR
free} \ &\ \\ \hline
$|$ & $\Downarrow$ & : & *  & \ \ \ \ \ \ \ \  \ \ \  * & : &\ $\Uparrow$ &\
\\
$\Nc+2$ & $\Downarrow$ & : & *  & \ \ \ \ \ \ \ \ \ \ \ * & : &\ $\Uparrow$ &\
\\
\  & $\Downarrow$ & ${11\over 9}\Nc$ & *  & {\bf IR}\ \ \ \ \ \ \
 * & ${11\over 2}\Nc$ &\
$\Uparrow$ &\ \\
\  & $\Downarrow$ & : & ${9\over 7}\Nc$  &  {\bf fixed }\ \  ${9\over
2}\Nc$ & : &\ $\Uparrow$ &\ \\
\ & $\Downarrow$ & : & *  & {\bf point} \ \ \   * & : &\ $\Uparrow$ &\
\\
$|$ & $\Downarrow$ & : & *  & \ \ \ \ \ \ \ \ \  \ \  * & : &\ $\Uparrow$ &\
\\
\hline
{magnetic  theory} & {\bf IR free} & : & *  & \ \ \ \ \ \ \ \ \ \
{\bf AS\ free}
&  &\  \ &\ \\ \end{tabular}
\end{center}

The picture describes a scenario
where in the range between the $:$ signs, namely
 ${11\over 9}N_c < N_f < {11\over 2}N_c$ for the $R$ case and
between the $*$ signs, namely,
${9\over 7}N_c < N_f < {9\over 2}N_c$ for the $\notR$ case, the
two theories flow to  a common fixed point in the IR.
In the range $\Nf<{11\over 9}N_c\ (\Nf<{9\over 7}N_c)$  the $R$ ($\notR$)
system flows to the  free magnetic theory in the IR, whereas for $\Nf>{11\over
2}N_c\ (\Nf>{9\over 2}N_c)$ the picture in the  infrared for  $R$ ($\notR$) is
that of the free electric theory.

The discussion of the decoupling limit for these theories is very
similar to that for the $\Nf = \Nc $ theory.
Thus,  for those values of $\Nf$ which have massless fermions for small
values of $m_Q$  we must have a second-order phase transition as $m_Q$ is
increased.  It is not clear how the theory behaves on the other side of
this phase transition. In the $\notR$ case obviously only a $SU(\Nf)\times
U(1)_B$ symmetry remains, with no massless fermions.
This situation is schematically described as follows
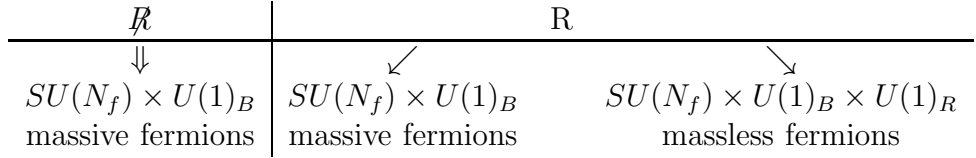
\begin{center}
\begin{tabular}{c | c c c }
$\notR$ & & R & \\ \hline
$\Downarrow$ & $\swarrow$ & & $\searrow$ \\
$SU(\Nf)\times U(1)_B$ &$SU(\Nf)\times U(1)_B$ & &
$SU(\Nf)\times U(1)_B \times U(1)_R$ \\
massive fermions & massive fermions & & massless fermions \\
\end{tabular}
\end{center}



\section{Problems of Approximate Supersymmetry on the Lattice}

(i) Can the phenomena we have discussed in this work be seen in lattice
gauge theory simulations?
  In general, gauge
theories on the lattice cannot be made supersymmetric at the
fundamental level. We expect that lattice simulations of these theories
will also contain small dimension 4 perturbations which violate
supersymmetry.  Our analysis has been based on the assumption that, if
the phenomena discussed by Seiberg survive perturbations which are
relevant in the infrared, they should also survive small marginal
perturbations.

(ii)   Can one reach the
continuum limit in such lattice simulations?
 Typically in lattice gauge theory simulations
with scalar fields, there is no continuum limit; instead, one finds a
first order phase transition as a function of the scalar field mass
parameter
\cite{CandWn}.
  Renormalization effects in a gauge theory can
induce an unstable potential for a scalar field coupled to the gauge
bosons, leading to a `fluctuation-induced first-order phase
transition'.  How can this be avoided?

(iii) To analyze this question, consider the renormalization group equations
for an approximately supersymmetric gauge theory.  Viewed as a
conventional renormalizable gauge theory, SQCD has three coupling
constants, the gauge coupling $g$, the quark-squark-gluino coupling
$g_\lambda$, and the four-scalar coupling $g_D$.  The scalar potential
has the  specific form
\beq
    V = { g_D^2\over 2} \biggl[ Q^\dagger \tau^A Q
    - \Qb \tau^A \Qb^\dagger
                \biggr]^2 \ ,
\eeq{Dpot}
where $\tau^A$ is an $SU(\Nc)$ matrix.
If we relax the constraint of supersymmetry, there are four possible
invariants under the symmetries of the problem, including the
continuous global symmetries and parity $Q \leftrightarrow
\Qb^\dagger$. The most general linear combination of these invariants
can be generated by the renormalization group flow.

We can view the effects of renormalization relatively simply, however,
by restricting our attention to the vicinity of the surface given by
the three couplings $g$, $g_\lambda$, and $g_D$.  In this surface, the
beta functions of the three couplings are given by
\beqa
   \beta_g &=& -{1\over (4\pi)^2}\bigl[ 3\Nc - \Nf \bigr] g^3 \CR
   \beta_{g_\lambda} &=& -{1\over (4\pi)^2}\biggl[ g_\lambda g^2
      \bigl( 3\Nc + 3 C_2(\Nc)\bigr)
            - g_\lambda^3 \bigl( 3C_2(\Nc) +
            \Nf\bigr)\biggr] \CR
   \beta_{g_D^2} &=& - {1\over (4\pi)^2} \biggl[ 4g_\lambda^4 \Nc
                         + 2 g_D^4 \bigl(\Nc - \Nf - 2C_2(\Nc)\bigr)
                   + 12 g_D^2 g^2 C_2(\Nc) - 8 g_D^2 g_\lambda^2
                   C_2(\Nc) \biggr] \ ,\nonumber
\eeqan
where $C_2(\Nc) = (\Nc^2-1)/2\Nc$.  These three functions all reduce to
the standard SQCD beta function on the supersymmetric subspace; for
 $g^2 = g_\lambda^2 = g_D^2$, $\beta_g = \beta_{g_\lambda} =
 \beta_{g^2_D}/2g$.  Note that, for $\Nc \sim \Nf $ and the three
    couplings in reasonable ratio, all three couplings are
    asymptotically free.

 The potential instability to a first order phase transition
 arises because a new structure in the potential is induced by the
 renormalization group flow. To lowest order, the form of the potential
 induced is
 \beq
    V_E = { g_E^2\over 2} \biggl[ Q^\dagger
    \{\tau^A,\tau^B\} Q
    + \Qb \{\tau^A,\tau^B\} \Qb^\dagger
                \biggr]^2 \ .
\eeq{Epot}
On the surface $g_E^2 = 0$, the beta function for $g_E^2$ is
 \beq
   \beta_{g_E^2} = - {1\over (4\pi)^2} \biggl[ 4g_\lambda^4
                              - 3 g^4 - g_D^4 \biggr] .
\eeq{Ebetas}
This equation implies that, if
one  leaves out the gluinos, $g_E^2$ becomes negative in the
 infrared, leading to  a fluctuation-induced first-order phase
 transition.  According to \leqn{Ebetas}, this effect is removed if the
 lattice simulation includes gluinos, and if the gluino coupling
 $g_\lambda$ is large enough.

 With this provision to avoid possible first-order phase transitions,
 we expect that lattice simulations with an approximately
 supersymmetric action can reach the continuum limit and test our
 predictions for softly broken supersymmetric QCD.

  \Acknowledgements
We are grateful to Eliezer Rabinovici and the organizers of the 1995
Jerusalem Winter School for bringing us together. We thank Nathan Seiberg
for a stimulating set of lectures at the school which ignited our
interest in this problem and for several useful discussions.
  We thank Shmuel Nussinov and Gabrielle Veneziano for
stimulating conversations, and we are especially grateful to Tom Banks for
emphasizing the importance of QCD inequalities.  MEP thanks
the members of the high energy group for their
hospitality at Tel Aviv University during the initial phase of this
work.

\end{document}